\documentclass[usenatbib,useAMS]{mn2e}
\usepackage{epsfig,psfig,graphicx,float,color}
\usepackage{subfigure}
\usepackage{mybibdefs}
\usepackage{multirow}
\usepackage{widetext}
\usepackage{url}
 \usepackage{color}
\usepackage{amsmath}

\newcommand\rr{\color{black}}
\newcommand\rrr{\color{black}}

\usepackage{graphicx}
\usepackage{amssymb}
\usepackage{epstopdf}
\title[Feature importance for machine learning redshifts]{Feature importance for machine learning redshifts applied to SDSS galaxies}  
\author[Hoyle et al.]{Ben  Hoyle$^{1,2}$, Markus Michael Rau$^{1}$, Roman Zitlau$^{1}$, Stella Seitz$^{1,3}$, Jochen Weller$^{1,2,3}$ \\\\\\\\
$^1$Universitaets-Sternwarte, Fakultaet fuer Physik, Ludwig-Maximilians Universitaet Muenchen, Scheinerstr. 1, D-81679 Muenchen, Germany\\
$^2$Excellence Cluster Universe, Boltzmannstr. 2, D-85748 Garching, Germany\\
$^3$Max Planck Institute for Extraterrestrial Physics, Giessenbachstr. 1, D-85748 Garching, Germany. \\
\\
{\tt E-mail: hoyleb@usm.uni-muenchen.de}
 }
  
\begin{document}
\date{Accepted ----. Received ----; in original form ----.}
\pagerange{\pageref{firstpage}--\pageref{lastpage}} \pubyear{2010}
\maketitle
\label{firstpage}
\begin{abstract}
We present an analysis of importance feature selection applied to photometric redshift estimation using the machine learning architecture  {\rrr  Decision Trees with the ensemble learning routine Adaboost (hereafter RDF)}. We select a list of 85 easily measured (or derived) photometric quantities (or `features') and spectroscopic redshifts for almost  two million galaxies from the Sloan Digital Sky Survey Data Release 10. After identifying which features have the most predictive power,  we use standard artificial Neural Networks (aNN) to show that the addition of these features,  in combination with the standard magnitudes and colours, improves the machine learning redshift estimate by 18\% and decreases the catastrophic outlier rate by 32\%. We further compare the redshift estimate {\rrr using  RDF} with those from two different aNNs, and with photometric redshifts available from the SDSS. We find that the RDF requires orders of magnitude less computation time than the aNNs to obtain a machine learning redshift while reducing both the catastrophic outlier rate by up to 43\%,  and the redshift error by up to 25\%. When compared to the SDSS photometric redshifts, the RDF machine learning redshifts both decreases the standard deviation of residuals  scaled by 1/(1+z)  by $36\%$ from 0.066 to 0.041, and decreases the fraction of catastrophic outliers by $57\%$ from 2.32\% to 0.99\%.
\end{abstract}

\section{introduction}
Large scale photometric galaxy surveys require precise redshift information to maximize information about cosmological parameters. Obtaining accurate spectroscopic {\rr redshifts is costly and time intensive} and is typically only performed for a small subsample of all galaxies. Conversely, the measurement of multi-band photometric properties of galaxies is much cheaper. The compromise is then to attempt to extract less accurate redshift information from photometrically measured properties, but applied to the full galaxy sample. This paper examines which photometric properties, or `features', of the galaxies are best suited to this task by using feature importance analysis within standard machine learning architectures.

Photometric redshifts are often derived from galaxy Spectral Energy Distribution (hereafter SED) templates. The template redshift approach is well studied and is physically {\rr motivated}. We know how the measured flux of a fiducial galaxy will change with redshift, and we can employ our knowledge of stellar populations and their evolution to predict how the SED and therefore fluxes and colors of galaxies will change as a function of redshift and galaxy type. However the encoding of the complex stellar physics, the computational time required to generate the templates, coupled with our uncertainty in the stellar population models and observed measurement error, combine to produce redshift estimates which are little better than non-parametric techniques \citep[see e.g.,][for an overview of different techniques]{2010A&A...523A..31H,2013ApJ...775...93D}.

Machine learning methods offer a non-parametric alternative to template methods. Generally the `machine architecture' learns how to {\rr combine and (or) weight and (or) cluster the photometric galaxy properties to produce a machine learning redshift}. The machine then examines the best combinations (and) or clustering (and) or weighting to minimize the difference between the spectroscopic redshift (of a training sample) and the machine learning redshift. 

A major advantage of the template method over the machine learning method is the need of the latter to have a well defined training set which spans the input feature parameter space of interest. On the other hand, one may view the need to generate templates and, importantly, a non-biased sample of templates, an equally large obstacle in order to produce reasonable template method redshifts. {\rr There has also been work to combine template and machine learning techniques \citep[e.g.,][]{2006MNRAS.372..565F,2009ApJ...690.1236I} so the  line of distinct has become less clear.}

Non parametric photometric redshift estimation techniques have been developing since \cite{1995AJ....110.2655C} and moved into the field of machine learning with the popular artificial Neural Network (aNN) code called aNNz \citep[][]{2004PASP..116..345C}. Since then a plethora of machine learning architectures, including Random Decision Trees, have been applied to the problem of redshift estimation \citep[see e.g.][for a list of routine comparisons]{2014arXiv1406.4407S}, and to estimate the full redshift probability distribution function \citep[][]{2010ApJ...715..823G,tpz}. Machine learning architectures have also had success in other fields of astronomy such as galaxy morphology identification, and star \& quasar separation \citep[see for example][]{1997daa..conf...43L,2009arXiv0910.3770Y}.

Most recently \cite{2014A&A...568A.126B} applied an advanced type of deep aNN to a subset of galaxies drawn from the Sloan Digital Sky Survey \citep[hereafter SDSS,][]{2000AJ....120.1579Y}  Data Release 9 \citep[][]{2012ApJS..203...21A}. The galaxy training sample was selected to be clean of artifacts, and to be confirmed spectroscopically as a galaxy. The resulting machine learning redshift error for this clean subset of galaxies is $\sigma_z=0.023$, several factors smaller than the photometric redshift available in the SDSS CasJobs interface for the same sample of galaxies.

The machine learning community uses the nomenclature `features' for items which are input into a machine architecture. For our purposes, the features can be any easily measured, or derived, photometric quantities that are available for each galaxy. For example standard machine learning redshift analysis uses a set of input features drawn from either fluxes or magnitudes, or a pivot magnitude and colors. 

However one has the freedom to choose other features which are easily measured photometrically and which may also scale with distance. For example it is conceivable that the observed galaxy size, or the iso-photal radius in some (or all) bands may also encode redshift information. One may also find that galaxy inclination, or galaxy type as measured by the SDSS `fracDev' parameter, or Stokes parameters, may also be valuable to identify (perhaps morphological) subsamples which may each have different redshift scalings \citep[see][]{2011ApJ...730...54Y}. Given the quantity of easily obtained photometric parameters, and their ease of accessibility, it seems pertinent to explore if the addition of extra features can indeed improve machine learning redshift estimates. 

Indeed early work using aNNs by \cite{2003LNCS.2859..226T} find that the inclusion of radii and fluxes in addition to magnitudes improves the machine learning redshift estimation compared with using just magnitudes. However \cite{2011PASP..123..615S} use more derived morphological input features and a principle component type analysis to show that the addition of the examined features do not drastically improve machine learning redshift estimation. Furthermore \cite{2006astro.ph.12749L} and \cite{2013ApJ...772..140B} explore different magnitude definitions as input features of aNNs, and find that some magnitudes produce more accurate machine learning redshift estimates than others (e.g. dereddened model magnitudes from the SDSS).

We expand these previous result by compiling 85 standard (e.g. magnitudes) and extended photometric features, and then use feature importance to determine which features have the most predictive power when estimating the galaxy redshift.  This is performed using standard feature selection analysis well known to the machine learning community. A full feature importance analysis with so many different features has yet to be applied to machine learning redshift estimation.  

We examine the use of the following machine learning architectures: {\rrr  decision trees combined using the Adaboost algorithm} to perform the feature importance and to measure redshifts, and standard aNNs to see how the effect of selecting different features changes the recovered machine learning redshifts. We note that some of the aforementioned machine learning architectures are extremely scalable, and can be performed in a matter of tens of seconds on today's desktop computers with sample sizes of millions.

{\rr The format of the paper follows. We describe the data sample and the list of measured and derived photometric features in \S\ref{data}. We continue by detailing the machine learning architectures applied in this work, and introduce the feature importance in \S\ref{method}. We describe the analysis and present results in \S\ref{results}, and conclude and discuss in \S\ref{conclusions}.}

\section{Data}
\label{data}

The data in this study are drawn from SDSS Data Release 10 \citep[][]{2014ApJS..211...17A}. The SDSS I-III uses a 4 meter telescope at Apache Point Observatory in New Mexico and has CCD wide field photometry in 5 bands \citep[$u,g,r,i,z$][]{Gunn:2006tw,Smith:2002pca}, and an expansive spectroscopic follow up program \citep[][]{2011AJ....142...72E} covering $\pi$ radians of the northern sky. {\rr The SDSS collaboration has obtained approximately 2 million galaxy} spectra using dual fiber-fed spectrographs. An automated photometric pipeline performed object classification to a magnitude of $r\approx$22 and {\rr measured} photometric properties of more than 100 million galaxies. The complete data sample, and many derived catalogs such as the photometric redshift estimates, are publicly available through the CasJobs server\citep[][]{10.1109/MCSE.2008.6}\footnote{skyserver.sdss3.org/CasJobs}.

The SDSS is well suited to the analysis presented in this paper due to the enormous number of photometrically selected galaxies with spectroscopic redshifts to use as training, cross-validation and test samples. We select 1,958,727 galaxies from CasJobs with both spectroscopic redshifts and photometric properties. In detail we run the following MySQL query in the DR10 schema:
{\rrr
\begin{verbatim}
SELECT s.specObjID, s.objid, s.ra,s.dec, 
p.z AS photoz, p.zerr AS photoz_err,
s.z AS specz, s.zerr AS specz_err,
s.dered_u,s.dered_g,s.dered_r,s.dered_i,
s.dered_z,s.modelMagErr_u,s.modelMagErr_g,
s.modelMagErr_r,s.modelMagErr_i,s.modelMagErr_z,
s.type as specType, q.type as photpType,
q.petroRad_u,q.petroRad_g,q.petroRad_r,
q.petroRad_i,q.petroRad_z,
q.petroRadErr_u,q.petroRadErr_g,q.petroRadErr_r,
q.petroRadErr_i,q.petroRadErr_z,
q.deVRad_u,q.deVRadErr_u,q.deVRad_g,q.deVRadErr_g,
q.deVRad_r,q.deVRadErr_r,
q.deVRad_i,q.deVRadErr_i,q.deVRad_z,q.deVRadErr_z,
q.extinction_u,q.extinction_g,q.extinction_r,
q.extinction_i,q.extinction_z,
q.psfMag_u,q.psfMagErr_u, 
q.psfMag_g,q.psfMagErr_g, 
q.psfMag_r,q.psfMagErr_r, 
q.psfMag_i,q.psfMagErr_i, 
q.psfMag_z,q.psfMagErr_z, 
q.fiberMag_u,q.fiberMagErr_u,
q.fiberMag_g,q.fiberMagErr_g,
q.fiberMag_r,q.fiberMagErr_r,
q.fiberMag_i,q.fiberMagErr_i,
q.fiberMag_z,q.fiberMagErr_z,
q.expAB_u,q.expRad_u,q.expPhi_u,
q.expAB_g,q.expRad_g,q.expPhi_g,
q.expAB_r,q.expRad_r,q.expPhi_r,
q.expAB_i,q.expRad_i,q.expPhi_i,
q.expAB_z,q.expRad_z,q.expPhi_z
    
INTO mydb.specPhotoDR10v2 FROM SpecPhotoAll AS s 

JOIN photoObjAll AS q 
ON s.objid=q.objid AND q.dered_u>0 
AND q.dered_g>0 AND q.dered_r>0 
AND q.dered_z>0 AND q.dered_i>0 
AND q.expAB_r >0

LEFT OUTER JOIN Photoz AS p ON s.objid=p.objid
\end{verbatim}
}
We apply the SDSS extinction corrections to the psf and fiber magnitudes, and further only select galaxies that have a photometric galaxy classification $type=3$, have spectroscopic redshifts, $r$ band magnitudes, and radii greater than zero. This reduces the sample size to 1,922,231 galaxies. 

\subsection{SDSS DR10 photometric redshifts}
 The SDSS photometric redshifts are generated using a hybrid technique of the template method \citep[][]{2000AJ....120.1588B} and a machine learning component using k-nearest neighbours \citep[][]{2007AN....328..852C} technique as described in \cite{2009ApJS..182..543A}. We hereafter refer to this combined method as `template-ml'. The SDSS template-ml photometric redshifts are available from within CasJob by using the above SQL query. %The bottom panel of Fig. \ref{template} shows a histogram of the redshift dispersion in five bins of spectroscopic redshift.

%\begin{figure}
%   \centering
% \includegraphics[scale=0.4, clip=true, trim=10 10 15  35]{SDSSsphoto_specz.pdf}
%   \caption{ \label{template} The SDSS spectroscopic redshift and photometric redshift from the template-ml hybrid approach. The top panel shows spectroscopic redshift against photometric redshift. The lower panel shows histograms of the redshift dispersion in five bins of spectroscopic redshift.}
%\end{figure}

\subsection{Input features}
\label{featscal}
Table \ref{inputFeatures} shows the list of photometric features used in this work. {\rr This is a large but non-exhaustive list of possible input features. There are still more photometric features one may choose to use, such as Petrosian magnitudes, other apertures, or more detailed surface profiles \citep[see e.g.,][]{2011PASP..123..615S}. }

These photometric features are drawn from the following categories. 
%We choose an extensive list of easily measured photometric features drawn from the following categories. 
{\rr Magnitudes:} corresponding to magnitudes measured in different bands and apertures, and color combinations created from them; Radii: measurements of sizes in different bands and with differing definitions; Morphology: how much of the light profile is best fit by one profile compared to others; and Shapes: ratio of major and minor ellipses measured in different bands and the Means Stokes parameters in each band. We list the full set of input features in Table \ref{inputFeatures} and note that their full description can be found on the Sky Server web page\footnote{skyserver.sdss3.org/public/en/help/browser/browser.aspx}. 

For each feature dimension we perform feature rescaling by subtracting the mean of the feature distribution and dividing by two times the standard deviation. Feature rescaling allows features with potentially vastly different scales to be given equal weight in the analysis. Throughout the remaining paper all references to features refer to these re-scaled features. 

\begin{table}
\begin{center}
  \begin{tabular}{ | c | c |} 
 Description & Feature name   \\ \hline
\multirow{6}{*}{ {\rr Magnitudes } } & dered\_u dered\_g dered\_r  \\
 & dered\_i  dered\_z \\
 & psfMag\_u psfMag\_g psfMag\_r \\
 & psfMag\_i psfMag\_z \\
 & fiberMag\_u fiberMag\_g fiberMag\_r \\
 & fiberMag\_i fiberMag\_z \\  \hline
 \multirow{5}{*}{Radii} & {\rr petroRad\_u petroRad\_g petroRad\_r } \\
 & petroRad\_i petroRad\_z \\
& expRad\_u expRad\_g expRad\_r \\
 & expRad\_i expRad\_z \\
 & deVRad\_u deVRad\_g deVRad\_r \\
 & deVRad\_i deVRad\_z  \\ \hline
 \multirow{15}{*}{Colors} & dered\_z-dered\_i dered\_z-dered\_r \\
 & dered\_z-dered\_g dered\_z-dered\_u \\
 & dered\_i-dered\_r dered\_i-dered\_g \\
 & dered\_i-dered\_u dered\_r-dered\_g \\
 & dered\_r-dered\_u dered\_g-dered\_u \\
 & fiberMag\_z-fiberMag\_i fiberMag\_z-fiberMag\_r \\
 & fiberMag\_z-fiberMag\_g fiberMag\_z-fiberMag\_u \\
 & fiberMag\_i-fiberMag\_r fiberMag\_i-fiberMag\_g \\
 & fiberMag\_i-fiberMag\_u fiberMag\_r-fiberMag\_g \\
 & fiberMag\_r-fiberMag\_u fiberMag\_g-fiberMag\_u \\
 & psfMag\_z-psfMag\_i psfMag\_z-psfMag\_r \\
 & psfMag\_z-psfMag\_g psfMag\_z-psfMag\_u \\
 & psfMag\_i-psfMag\_r psfMag\_i-psfMag\_g \\
 & psfMag\_i-psfMag\_u psfMag\_r-psfMag\_g \\
 & psfMag\_r-psfMag\_u psfMag\_g-psfMag\_u \\ \hline
 \multirow{2}{*}{Profile} & fracDeV\_u fracDeV\_g fracDeV\_r \\
 & fracDeV\_i fracDeV\_z \\ \hline
 \multirow{3}{*}{Ellipticity} & expAB\_u expAB\_g expAB\_r \\
 & expAB\_i expAB\_z \\
 & deVAB\_u deVAB\_g deVAB\_r \\
 & deVAB\_i deVAB\_z \\ \hline
 \multirow{3}{*}{Means Stokes} & q\_u u\_u q\_g u\_g \\
 & q\_r u\_r q\_i u\_i \\
 & q\_z u\_z\\ \hline
  \end{tabular}
\caption{\label{inputFeatures} The complete list of the input photometric features used in this work. The Means Stokes parameters are shape features. A full description of each of the parameters can be found on the SDSS Sky Server web page.}
\end{center}
\end{table}
The aim of this work is to identify which of these features provides the most predictive power for redshift estimation. In order to select these features, we perform feature importance, described in Section \ref{featimp}. We then obtain a ranking of features, in which the features which have the highest rank have the most predictive power and are selected as inputs into the final model.

\subsection{Training, cross-validation and test subsets}
We follow traditional machine learning nomenclature and methodology by randomly sub-dividing the galaxy catalog into a training, cross-validation and test samples with proportions $50\%,25\%,25\%$ respectively. The training sample is used to train the machine learning system for a given architecture and hyper-parameter set. One uses the cross-validation sample to select the best values for the hyper-parameters of the learned system. Once the set of hyper-parameters has been decided upon, neither the training nor cross-validation sample provide a bias free estimate of the true error. In these cases the test sample is used to measure the true ability of the learned machine to generalize to a new dataset.

\section{Machine learning methods}
\label{method}
Below we describe the two artificial Neural Network programs  used in this work and the machine learning frameworks Decision Trees with the ensemble learning code Adaboost.

\subsection{Artificial Neural Networks}
We use two different artificial Neural Network (aNN) architectures. The first is the Java based applet Photoraptor \citep[][]{2014PASP..126..783B,2014arXiv1406.3192C} and the second is the Fortran code FaNN \citep[][]{nissen03} which is callable from the command line and has wrappers in many languages including Python. 

Photoraptor is a standard multilayer perceptron with up to two (feed forward) hidden layers which trains the connections between neurons efficiently using a Quasi Newton Algorithm. We follow \cite{2014A&A...568A.126B} and use an architecture of two hidden layers of size (11,4). 

FaNN extends standard aNN architecture by incrementally building hidden layers and determining connections using the Cascade2 training algorithm \citep{Fahlman:1990:CLA:109230.107380}. Cascade2 training starts with an empty set of hidden layers and incrementally trains and adds one (multiply connected) neuron until either a user-specified maximum number have been added, or until the training error reaches some user-specified threshold.

The hyper-parameters of the aNNs explored in this work are the number of training examples (the training sample size), and the number of input features per training example, the number of hidden layers and neurons, and the method to learn the best connections between neurons.

\subsection{Decision Trees}
We use the Python package scikit-learn \citep[][]{scikit-learn} and the included implementation of {\rrr Decision Trees for regression} \citep[][]{ig}. There exist other public implementations of {\rrr trees and forests} for Classification and Regression, some of which estimate the full shape of the machine learning redshift probability distribution function \citep[e.g.][]{tpz}.

The Decision Tree Regressor sub-divides (or branches) the $N$ data with respect to the feature space $\mathbf{f}_{i}$ into $\tau$ branches $\mathcal{B}_{\tau}$ which end in $l$ leaf nodes per tree. The branches are constructed such that the spectroscopic redshift $z_{{\rm spec}, i}$ in each leaf has a low mean squared error 
\begin{equation}
MSE = \frac{1}{N} \sum_{\tau = 1}^{l} \sum_{\mathbf{f}_i \in \mathcal{B}_{\tau}} \left(z_{{\rm spec}, i} - 
\left\langle z_{{\rm spec}, \tau} \right\rangle \right)^2\,,
\label{eq:mse_tree}
\end{equation}
with respect to the mean spectroscopic redshift
\begin{equation}
  \left\langle z_{{\rm spec}, \tau} \right\rangle = \frac{1}{N_{\tau}} \sum_{\mathbf{f}_{i} \in \mathcal{B}_{\tau}} z_{{\rm spec}, i} \thinspace .
\label{eq:pred_reg_tree}
\end{equation}
Where $N_{\tau}$ is the number of objects in each leaf, The branches of the tree correspond to regions of input feature space. The machine learning redshift of a new object is obtained by assigning the value of Eq. \ref{eq:pred_reg_tree} to the final leaf that the data falls upon. The tree is grown recursively such that the mean spectroscopic redshift in each leaf is similar to the spectroscopic redshift of the objects in the leaf. We measure this similarity using the mean squared error function (Equ \ref{eq:mse_tree}). The selection of the best input feature to split on, and the best splitting point, is determined using an exhaustive search to minimize Eq. \ref{eq:mse_tree}. 
The tree is grown until each leaf node contains $N_{\tau}$ objects, where $N_{\tau}$ is a hyper-parameter of the model.  We then determine the feature importance score by summing (and normalizing) the decrease in Eq. \ref{eq:mse_tree} for each feature. 

%Random Decision Trees sub-divides (or branches) the data in each feature dimension by determining the (approximate) median value. The branched regions are further sub-divided using the same algorithm until each leaf contains only one datum (i.e. is pure) or contains a user specified number of data. The selection of which dimensional to begin the branching process is random, and therefore points on one leaf after one complete training iteration may land on a different leaf (and that leaf may have a different shape) after a different complete training iteration. Each leaf combines the interesting properties of the training examples which are located on it.

Computing a {\rrr single Decision Tree} is incredibly fast. This implementation can partition one million galaxies along 85 feature dimensions in a few tens of seconds using only one CPU core.  Predicting a redshift for new data is also very fast, requiring just seconds on a dataset of size half of a million.

\subsection{Adaboost}
The power of the {\rrr Decision Trees} can be enhanced by combining many trees together to make a  {\rrr forest. We select randomly from input data to produce each forest so use the terminology RDF throughout the paper.} One method to achieve this is by using an {\rrr algorithm} called Adaptive Boosting or Adaboost \citep[][]{Freund1997119,Drucker:1997:IRU:645526.657132}. {\rrr In this work we define the collection of trees constructed using Adaboost as a forest. This should not be confused with normal random forests in machine learning, which build trees simultaneously.}
%In this work, Adaboost combines many random decision trees with different hyper-parameters by weighting each tree's accuracy when predicting the machine learning redshift. The weighting process continues by building similar Random trees again using the previous weights and updating those weights based on performance. The final redshift is then a weighted average of Random trees in the forest. The hyper-parameters of the RDF and the learning process used in this work are the number of trees and their randomized seeds, the minimum number of training objects on each leaf node (which is a high dimensional rectangle), and the number of training examples.
We give a brief overview of the algorithm below, and refer the reader to \citet{Drucker:1997:IRU:645526.657132} for a more detailed description of the algorithm used in the scikit-learn routine. We note that \cite{2010ApJ...715..823G} also use {\rrr Adaboost and trees}, but they determine the full shape of the machine redshift probability distribution function using SDSS magnitudes as input features.

The basic idea of boosting is to improve the performance of a base learner, in our case the Decision Tree Regressor, by using multiple models which put more weight on elements in the training set which have large prediction errors. 

The algorithm described in \citet{Drucker:1997:IRU:645526.657132} first assigns equal weight $w_i = 1$ to each galaxy in the training set. Subsequently one trains 
a Decision Tree Regressor on a new training set of size N, by bootstrap selecting N samples with replacement from the original training set. Each element has a probability to be selected given by
\begin{equation}
p_i = \frac{w_i}{\sum_{i = 1}^{N} w_i}\, .
\label{eq:prob_weights}
\end{equation}
This produces a new model which is added to the ensemble of models. The training set loss $L_{i}$, for each element $i$ is calculated as 
\begin{equation}
  L_{i} = \frac{|z_{\rm phot}(\mathbf{f}_i) - z_{{\rm spec}, i}|}{\underset{j}{\sup}|z_{\rm phot}(\mathbf{f}_j) - z_{{\rm spec}, j}|}\, ,
\end{equation}
where $z_{\rm phot}(\mathbf{f}_j)$ is the function represented by the corresponding tree. Note that $L_{i}$ is normalized in such a way, that $L_i \in [0, 1]$. 
We can then calculate the average loss $\overline{L}$ of the model using
\begin{equation}
  \overline{L} = \sum_{i = 1}^{N} L_{i} p_{i} \, ,
\end{equation}
where the sum runs over all elements in the training set.
The confidence $\beta$ for a specific model is defined by
\begin{equation}
  \beta = \frac{\overline{L}}{1 - \overline{L}}  \, ,
\end{equation}
and the weights for each model are iteratively updated by multiplying the weights for each element in the 
training set by $\beta^{1 - L_{i}}$. 

The weight update procedure gives less weight to elements with a low prediction error $L_{i}$ and therefore these objects are less likely to be included in the training set drawn in the next boosting iteration.
This focuses subsequent learners on elements with a high prediction error \citep[][]{Freund1997119,Drucker:1997:IRU:645526.657132}.
We train a number of Decision Tree Regressors in this way and update the 
weights for the training set. The number of trees $M$ included into the ensemble is a hyper-parameter of the model. 
If we query a new object with input features $\mathbf{f}_i$, we obtain a prediction
$z_{{\rm phot}, j}(\mathbf{f}_i)$ for each tree in the ensemble $j \in {1, \dots, M}$. 
The final machine learning redshift prediction $z_{\rm phot}(\mathbf{f}_i)$ is then given as the weighted median of the redshift predictions of the
models in the ensemble with respect to $\log{\left(1\big/\beta_j\right)}$ \citep[as described in][]{Drucker:1997:IRU:645526.657132}.

\subsection{Feature importance selection}
\label{featimp}
One noteworthy feature of RDF is the ability to determine which of the feature dimensions encode the most information about the quantity of interest, which here is the machine learning redshift. {\rr The scikit-learn implementation of {\rrr Decision Trees} uses the Gini importance (described below) to determine the predictive power of each feature. In detail, the Adaboost routine combines the feature importances for each tree to create a forest-wide feature importance following this procedure. First the feature importance of each {\rrr Decision Tree} is determined and then the same weight is applied to each feature importance value as applied to the tree when constructing the {\rrr forest}. The final output value of the feature importance provided by Adaboost is the sum of the individual tree importances normalized by the sum of the weights applied to each tree. }

The Gini importance is constructed from the Gini coefficient, which is the value of the MSE (Eq. \ref{eq:mse_tree}) of the items on each branch. The larger the MSE the larger the Gini coefficient. As more sub branches are formed the data on each sub branch become more homogeneous which reduces the Gini coefficient. The Gini importance measures the reduction in the Gini coefficient from the parent branch to the child sub branches. For our purposes, the higher the Gini importance, the more  the feature is able to separate the training data into similar redshift groups, and therefore the more predictive power the feature has. 

In summary, the more branches in the {\rrr Decision Tree} that a particular feature dimension has, the more predictive power it has when estimating the redshift compared with other feature dimensions.

\section{Analysis and Results}
\label{results}
We first determine which are the most important features using {\rrr RDFs}. We then document how the standard artificial Neural Network machine learning redshift is improved by the addition of the most important features. We then present what effect the size of the training sample has on the recovered machine learning redshift and finally show the further improvement in the photometric redshift of SDSS galaxies when using RDFs.

\subsection{Feature importance}
\label{feature_importance}

{\rr To determine feature importance we construct 25 {\rrr forests}, and for each forest we vary the following hyper-parameters: The number of trees, the number of objects on each leaf node, the random seed, and the size of the training set.} Upon completion of each forest Adaboost returns the list of input features with their representative importance weights. We rank the features by weight and extract the top 1,2,3 features for each forest. We construct a results cube which lists how often each feature has made it to a given importance rank. 

\begin{figure}
   \centering
 \includegraphics[scale=0.48, clip=true, trim=8 35 45 60]{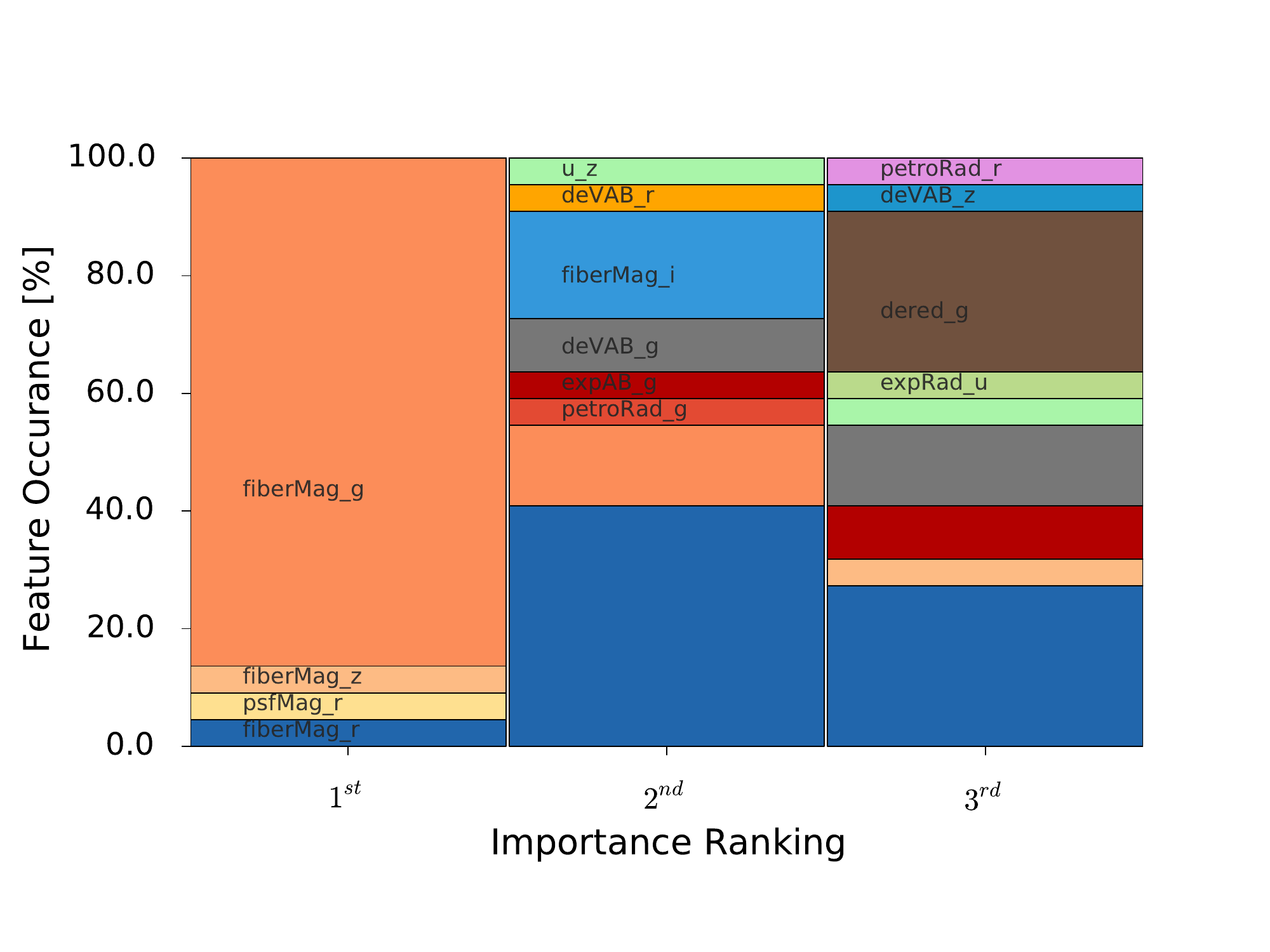}
   \caption{ \label{rankFeat} The relative importance of different photometric features when determining a machine learning redshift using SDSS data. The first column shows the most important, or top ranked, feature after each iteration. The second column shows the second most important feature. The height of the color bar is the occurrence rate that the feature has been deemed to be the most important. Note that for each forest we randomly select the machine learning architecture hyper-parameters and training sample size. The x-axis labels corresponds to the ranked importance of the features and we label each feature. The colors represent the same feature across each column.} 
\end{figure}

In  Fig. \ref{rankFeat} we show the relative importance of different photometric features when determining a machine learning redshift. The first column shows the most important, or top ranked, feature after each iteration. The second column shows the second most important feature. The height of the color bar is the occurrence rate that the feature has been deemed to be the most important. {\rr Note that for each forest we randomly select the machine learning architecture hyper-parameters and the training sample size.} The x-axis labels corresponds to the ranked importance of the features and we label each feature on the figure. The colors represent the same feature across each column.

Surprisingly, we find that the most important feature is the value of the fiber magnitude measured in the $g$ band, which, to the authors knowledge, has never been used in machine learning frameworks. We note that the SDSS fiber magnitude is a measure of the flux within a small (2 arc sec) radius around the galaxy center and is measured for all galaxies, not only the spectroscopic subset. We also note that the fibermag\_g is not only the most important feature, but it is often more important by a factor of 3 more than the second most important feature. 

{\rrr 
Given that we are using spectroscopic galaxies and selecting only galaxies with clean spectra, this implies that only galaxies whose flux is well defined within the fiber radius make the finally selection. Therefore these galaxies fiber apertures are probably a reasonable approximation of the underlying galaxy SED, and thus the apparent magnitude will scale with distance. 
}

To improve statistical significance we generate a further 350 further forests, and repeat the feature analysis. We find fiberMag\_g is still the most important, top 1, feature 67\% of the time. However 25 other features also appear at least once in the top 1 feature ranking, and we decide to not include these in Fig. \ref{rankFeat} to improve readability.

\begin{figure}
 \includegraphics[scale=0.45,clip=true,trim=15 35 45 70 ]{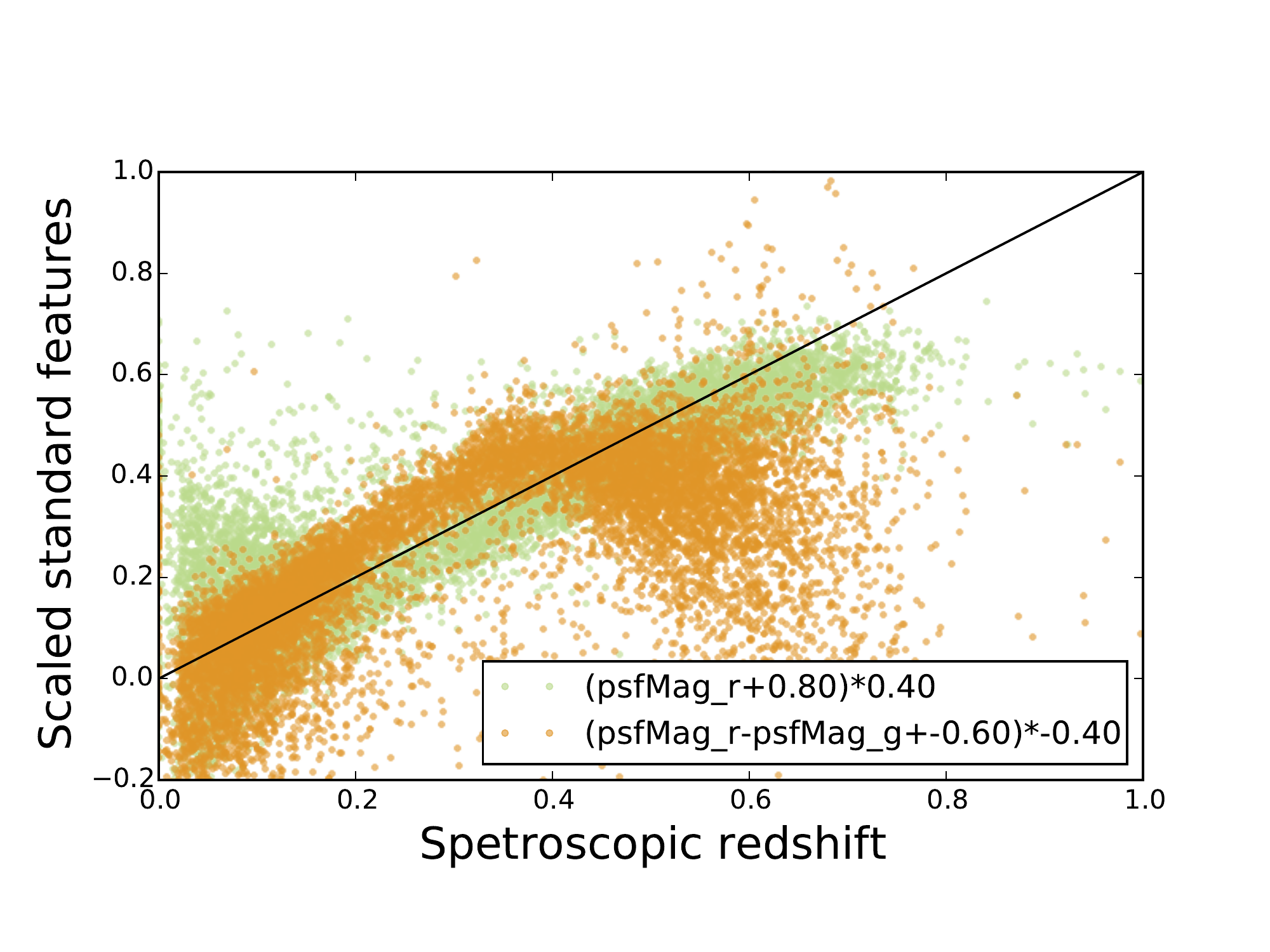}
 \includegraphics[scale=0.45, clip=true, trim=15 35 45 70]{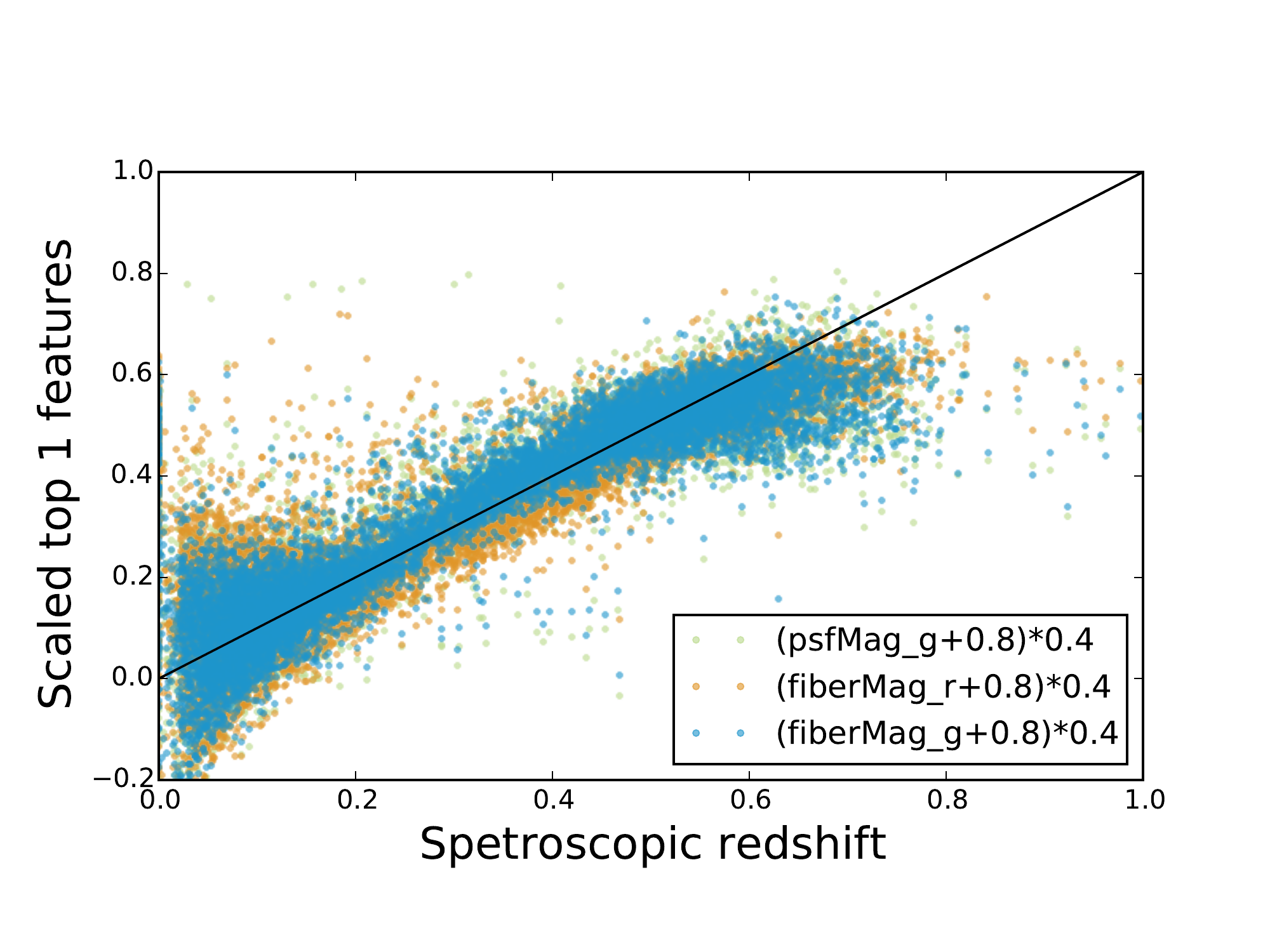}
 \caption{ \label{f2} An illustrative example of the power of the most important features at determining galaxy redshift. The top panel shows two standard features used in machine learning \citep[][]{2014A&A...568A.126B}. The bottom panel shows the three most important features as determined by this work. The x-axis is the true redshift and the y-axis shows the arbitrarily scaled features. The same sample of galaxies is shown in both plots. Note that all features are pre-scaled by  subtracting the mean and dividing by twice the standard deviation, see \S\ref{featscal}.}
\end{figure}

{\rr As an illustrative example} we examine how the top features scale as a function of redshift and show this illustrative plot in Fig. \ref{f2}. Recall that each of the input features are pre-scaled by subtracting the mean and dividing by twice the standard deviation, see \S\ref{featscal}. We have arbitrarily re-scaled the features, in a manner such as some of the machine architectures may use. Note that this is an illustrative explanation of the power of the most important features at determining galaxy redshift. The top panel shows two of the standard features used in machine learning \citep[see e.g.,][]{2014A&A...568A.126B}. The bottom panel uses the top three most important features as determined by this work. The data in this illustrative example is drawn from the test sample, which has not been used during machine training, and is therefore independent of the analysis. The same sample of galaxies is shown in both plots.

{\rr We also remove the fiber magnitudes and fiber colors completely from the analysis, and repeat the feature importance as above. We find that the dered\_g model magnitude is now the most important feature $70\%$ of the time. }

\subsection{The effect of the extended features}
\label{additional_features}
In this section we show what effect the use of extended features has on the machine learning redshift error estimated using popular frameworks. We choose two types of artificial Neural Networks, and fix their architectures in one approach, and allow it to vary in the other. 

In the first approach we use Photoraptor and fix the size of the training and test sets to be 120139, 360417, and the number of hidden layers to 2 and hidden units to (11,4). We note that here, the sum of the training and cross-validation data set corresponds to 25\% of the full data sample. Following \cite{2014A&A...568A.126B} we refer to the following features as  `standard': 
psfMag\_g-psfMag\_u,psfMag\_r-psfMag\_g,psfMag\_i-psfMag\_r,psfMag\_z-psfMag\_i,psfMag\_r. We use the results from \S \ref{feature_importance} to determine which are the top 1 and 2, and then top 1, 2, and 3, most important `extended'  features. The extended features are labeled in Fig.\ref{rankFeat}. We note that the top 1 and 2 features have dimension 10, and the top 1, 2 and 3 have dimension 14. 

We construct data vectors corresponding to the residual $\Delta_z$, between the true redshift $z$ and the machine learning redshift, and also these residuals scaled by $1/(1+z)$. On these vectors we compute and compare measurements of the mean, standard deviation, and the percentage of catastrophic `outliers` defined by $|\Delta_z/(1+z)|>0.15$. The measurements are performed on the following samples: the standard features, the top 1, and 2 best features, the standard \& the top 1,2 best features, and the standard \& the top 1, 2, and 3 best features. We show the results of this analysis using Photoraptor in Table. \ref{photoRapter}. 
\begin{table}
\begin{center}
  \begin{tabular}{ | c | c | c|} 
Input features & Measurement & Value \\ \hline
\multirow{3}{*}{Standard}&$\mu_{\Delta_z}\pm\sigma_{\Delta_z}$&-0.0001$\pm$0.075\\
&$\mu_{\Delta_z/(1+z)} \pm\sigma_{\Delta_z/(1+z)}$&-0.003$\pm$0.055\\
&$|\Delta_z/(1+z)|>0.15$&1.74\%\\ \hline
\multirow{3}{*}{Top1\&2}&$\mu_{\Delta_z}\pm\sigma_{\Delta_z}$&0.0001$\pm$0.068\\
&$\mu_{\Delta_z/(1+z)} \pm\sigma_{\Delta_z/(1+z)} $&-0.002$\pm$0.049\\
&$|\Delta_z/(1+z)|>0.15$&1.34\%\\ \hline
\multirow{3}{*}{Standard\&top1\&2}&$\mu_{\Delta_z}\pm\sigma_{\Delta_z}$ &-0.0001$\pm$0.066\\
&$\mu_{\Delta_z/(1+z)} \pm\sigma_{\Delta_z/(1+z)}$ &-0.002$\pm$0.046\\
&$|\Delta_z/(1+z)|>0.15$&1.22\%\\ \hline
\multirow{3}{*}{Standard\&top1\&2\&3}&$\mu\pm\sigma$&0.0001$\pm$0.065\\
&$\mu_{\Delta_z/(1+z)} \pm\sigma_{\Delta_z/(1+z)}$&-0.002$\pm$0.045\\
&$|\Delta_z/(1+z)|>0.15$&1.17\%\\
\\\hline
  \end{tabular}
\caption{\label{photoRapter} The values of the mean $\mu$, standard deviation $\sigma$, and the percentage of outliers for the residuals $\Delta_z$ between the true redshift and the Photoraptor machine learning redshift. We also show the results for the residuals scaled by $1/(1+z)$. We show the measured values for different input features sets (see text) and fix the training sample size and machine architecture hyper-parameters. Standard features are the psf magnitudes and colors, and the top 1,2,3 features are those labeled in Fig. \ref{rankFeat}. }
\end{center}
\end{table}

We find that the values of the mean, standard deviations and percentage of outliers decreases if we choose to use those features deemed to be the most important from \S\ref{feature_importance}. Furthermore we find that the combination of the standard \& top 1,2 (or standard \& top 1,2,3) features continue to improve each of the measured values. We find an improvement in the standard deviation of the redshift scaled residuals by $18\%$ and an improvement in the catastrophic outlier rate of 32\%. We note that the total training time of Photoraptor is approximately 4 hours for the stated hyper-parameters, on a single CPU core. The improvement in redshift estimation by combining radii, fluxes and magnitudes has been seen before \citep[][]{2010ApJ...715..823G}, however our choice of which additional features to use is motivated by the importance feature selection. 

In the second approach we allow the aNN architecture to vary. We calculate the machine learning redshift using the Cascade2 algorithm implemented in FaNN. We choose to examine the standard feature sample and the top 1\&2 feature sample. For this analysis we marginalize over the machine architecture hyper-parameters by randomizing the number of neurons in the hidden (multiply connected) layer, the learning rates, the desired best error the learning algorithm will try to attain, and the number of galaxies in the training set. We train FaNN using the training set, and calculate the standard deviation of the residuals using the full {\rrr test} set. 

We find that the value of the standard deviation of the residuals decreases by $17\pm8\%$ using the top 1\&2 features compared to the standard features. However the value of the standard deviation for the best hyper-parameter configuration is 0.081 which is not competitive with that obtained using Photoraptor.

Finally, as a further illustrative example of feature importance we determine which are the two worst, or lowest ranking, features from each iteration. The worst features are psfMag\_i-psfMag\_u, fracDeV\_g, fiberMag\_i-fiberMag\_u, fracDeV\_i, dered\_z-dered\_u, psfMag\_r-psfMag\_u, fracDeV\_r, fracDeV\_u, dered\_i-dered\_u, fracDeV\_z. We repeat the above analysis and pass these features to Photoraptor to measure a machine learning redshift. The value of the standard deviation of the residuals is 0.089 which is larger than using the standard, or top ranked features. We note that the outlier rate is also very high, with a value of 4.5\%. 

{\rr We present the following hypothesis which may describe why we could expect the least important features to be those listed above.
Features or feature combinations involving the u-band: Above a redshift of $0.1$ the 4000 Angstrom break drops out of the SDSS u-band. Red galaxies will no longer have a large measurement in this band which may indicate the lack of predictive power for redshifts. FracDev: This parameters describe morphology, e.g. how `bulge-like' or `disk-like' the galaxy is. We note that Singal et al. (2011) also show that the addition of some other morphological parameters such as concentration, does not improve the machine learning redshift.  }

\subsection{The importance of the size of the training set}
\label{ml_trainsize}
In this section we use the computationally fast {\rrr RDFs} as the machine learning architecture. This allows many different training sample sizes to be examined, and the other hyper-parameters to be explored in a timely manner. We sample the hyper-parameter space and generate {\rrr 1100 unique RDFs}.

\begin{figure}
   \centering
 \includegraphics[scale=0.45,clip=true,trim=0 0 50 35 ]{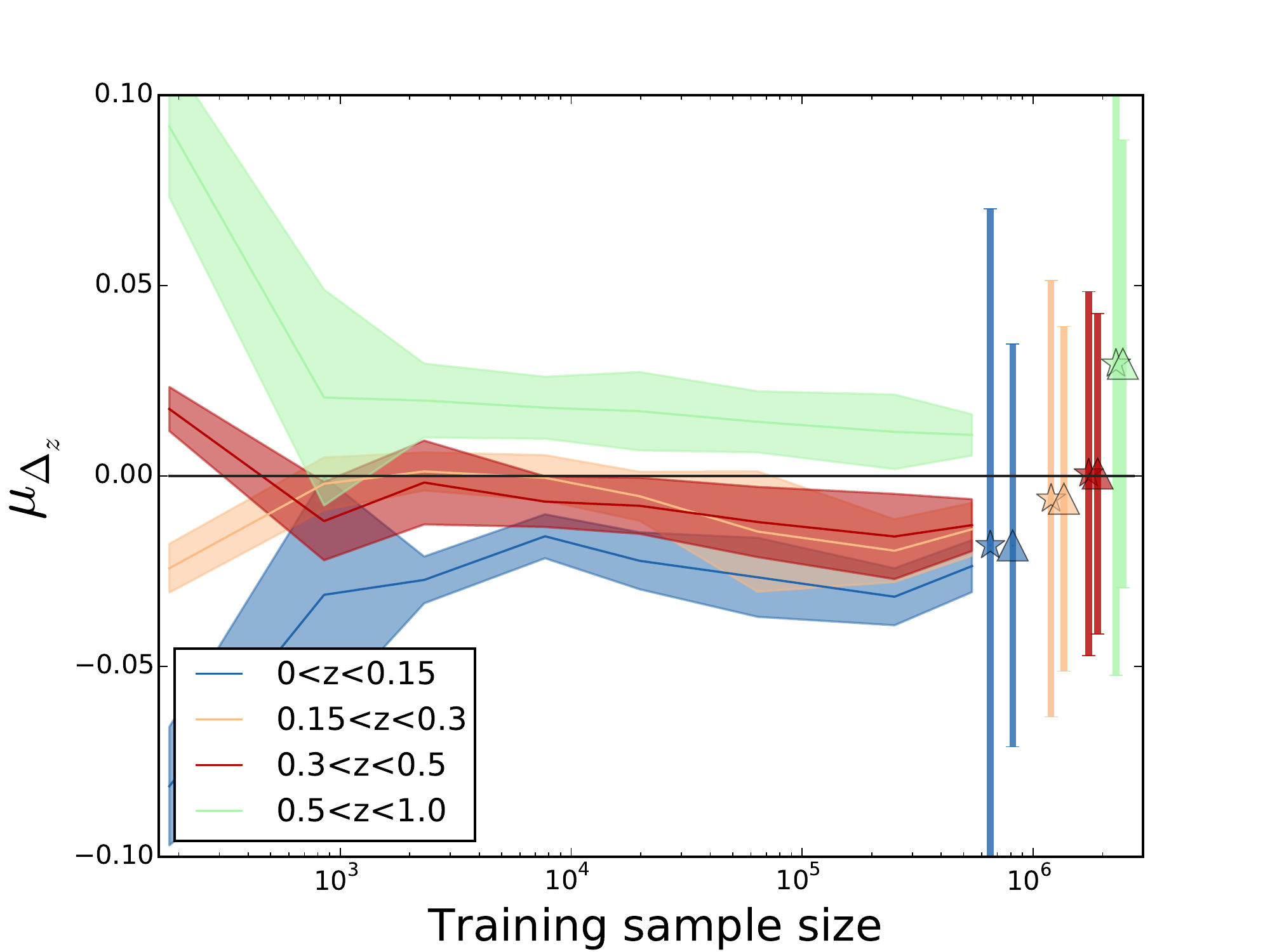}
   \caption{ \label{trainSize} The effect of the size of the training sample on the mean $\mu$ value of the redshift residuals $\Delta_z$, calculated using {\rrr forests}. The sample is divided into four redshift ranges (see legend) and the mean, error on the mean and the standard deviation is calculated for the residuals in each bin. The  central lines show the mean values, and the shaded regions show the error on the mean. The other hyper-parameters of the machine architecture are marginalized over. The large star symbols and thick error bars show the  mean and standard deviation values of the residuals calculated using the SDSS template-ml redshift calculated on the same cross-validation set. The large triangles show the smallest standard deviation of the machine learning method. We have positioned the triangles next to the stars to aid comparison. 
   }
\end{figure}

In Fig. \ref{trainSize} we show the effect of the size of the randomly selected training sample on the cross-validation mean $\mu$, and standard deviation of the residuals $\Delta_z$, in four redshift ranges (see legend).  The central lines show the mean values, and the shaded regions show the error on the mean. The large star symbols and thick error bars show the mean and standard deviation values of the residuals calculated using the SDSS template-ml redshift on the same cross-validation set. The large triangles show the smallest standard deviation of the machine learning method. We have positioned the triangles next to the stars to aid comparison. 

Fig. \ref{trainSize} shows that the bias $\mu_{\Delta_z}$ on the cross-validation sample decreases as the randomly selected training sample increases until the number is $\approx$2000 after which it decreases less significantly and approaches a constant. The mean values are well within one standard deviation of the line $\mu_{\Delta_z}=0$.  The dispersion of $\mu_{\Delta_z}$ at fixed training sample size is due to the random nature of the {\rrr RDF}, and also the marginalization over the RDF hyper-parameters. At fixed training sample size, we find that the mean and standard deviation in each redshift bin is largely unaffected by our choices of machine architecture hyper-parameters. We find that including more than $\sim$ 100,000 galaxies in the training set does not drastically improve the machine learning redshift when using the RDF method. Recall that the cross-validation sample is not shown to the machine architecture while training, and therefore presents an unbiased estimate of the error.

\subsection{The effect of the machine learning framework}
\label{ml_framework}
{\rrr RDFs} are constructed in \S\ref{feature_importance} to measure feature importance, but they also provide a precise measurement of the machine learning redshift, as previously seen by e.g., \cite{2010ApJ...715..823G,tpz}. The hyper-parameter values of the RDF are the number of trees and the minimum number of training objects on each leaf node. We fix the size of the input features to be all 85 features seen in Table \ref{inputFeatures}. We randomly explore the remaining hyper-parameter space. For each random instances of hyper-parameters the training sample (of size up to 961114) is used to train the RDF. For each instance the residuals are computed on the full cross-validation sample (consisting of a different set of 480557 galaxies). 

We choose the RDF with the lowest standard deviations as our final set of hyper-parameters, and finally measure the standard deviations on the residuals calculated using the test set (again, another different set of galaxies of size 480557). We reiterate that the cross-validation data set is not used during training, and the test data set is only used in this final stage. This results in an unbiased estimate of the error when applied to new data. 

\begin{figure}
   \centering
 \includegraphics[scale=0.45, clip=true, trim=20 2 10 35]{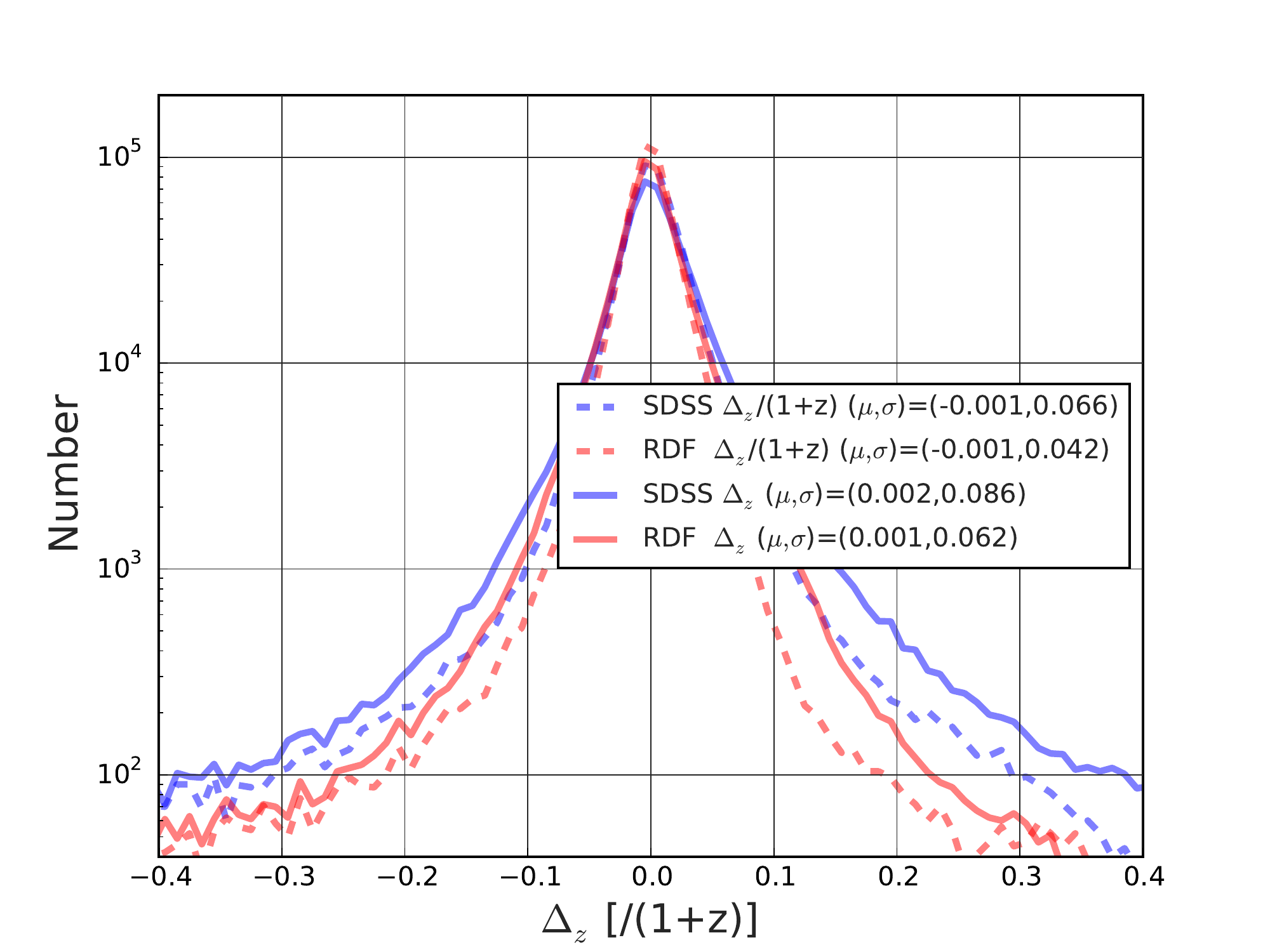}\\
 \includegraphics[scale=0.45, clip=true, trim=20 15 10 45]{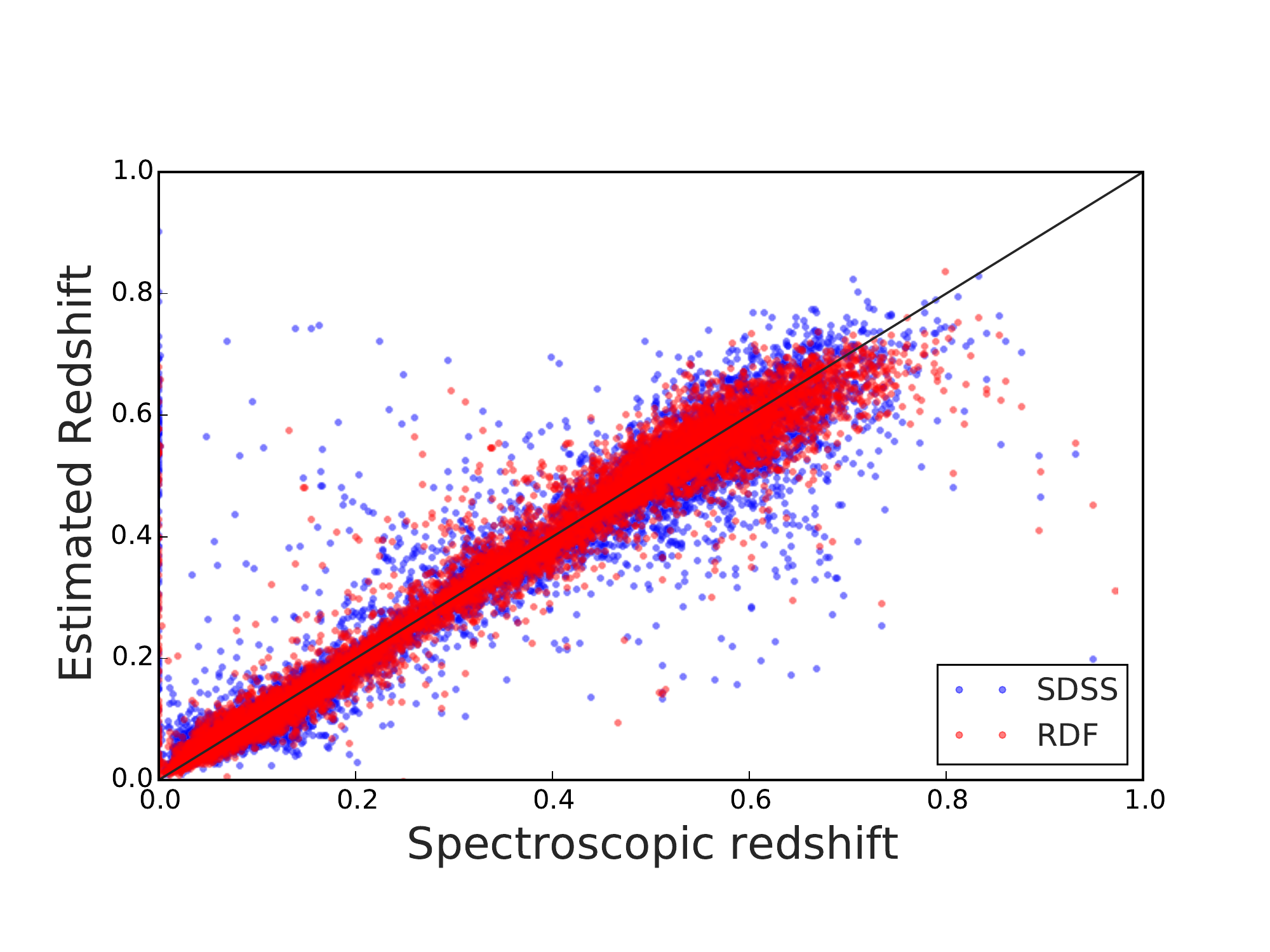}
   \caption{ \label{forstvsdr10} Top panel: The (blue) red curve shows the histogram of residuals between the true redshift and the (SDSS DR10 template-ml) {\rrr RDF}, redshift for the $480557$ galaxies in the test sample. The scale is logarithmic along the y-axis. The dotted lines show the residual distributions scaled by 1/(1+spec\_z). The legend shows the values of the mean $\mu$ and standard deviation $\sigma$ of each distribution. The test sample is not used during the RDF training or hyper-parameter selection. The hyper-parameter values of the RDF can be found in the text. Bottom panel: A scatter plot showing 10,000 randomly selected galaxies from the test sample. We show the spectroscopic redshift against the SDSS template-ml redshift and against the RDF machine learning redshift.
   }
\end{figure}

The top panel of Fig. \ref{forstvsdr10} shows the distribution of the test sample residuals between the true redshift and both the SDSS template-ml redshift, and the RDF redshift. We additionally show this distribution scaled by 1/(1+spec\_z) by the dotted lines, and mark the mean and standard deviation of the residual distribution for each case in the legend. The bottom panel of Fig. \ref{forstvsdr10} shows a redshift scatter plot using 10,000 randomly selected galaxies from the test sample. We show the spectroscopic redshift against the SDSS template-ml redshift and against the RDF machine learning redshift.

Fig. \ref{forstvsdr10} shows that the distribution of residuals using the RDF method (red lines) with 85 input features is more peaked around the true redshift, and declines faster as the residuals increase, than the distribution for the same sample of galaxies using the SDSS DR10 template-ml photometric redshift (blue lines). 

We find that mean and standard deviation of the residuals using the RDF machine learning redshift applied to the test sample to be $-0.001$ and $0.061$, and for the residuals scaled by 1/(1+spec\_z) we measure $-0.003$ and $0.041$. We next calculate the mean and standard deviation of the residuals on the test set using the SDSS template-ml photometric redshift and find them to be $0.002$ and $0.086$, and for the residuals scaled by 1/(1+spec\_z), we measure $-0.001$ and $0.066$. These values are also shown in Fig. \ref{forstvsdr10}. We note that the mean and standard deviations of the residuals of the RDF machine learning redshift are comparable, but slightly better than those measured using the artificial Neural Networks in Table \ref{photoRapter}. We reiterate that the RDF has used the full set of 85 input features, whereas the aNN were restricted to the standard, or top 1,2,3 most important features. {\rr A more complete analysis of the ability} of an aNN to measure a machine learning redshift with a much larger input features is planned for future work. 

We next restrict the input features of the RDF to be the `standard' features defined above, we measure the mean and standard deviation of the residuals $\Delta_z$ to be -0.002 and 0.075. This results in a slight improvement over the SDSS template-ml method, but is not as powerful as using the full 85 input features. We find that the outlier rate using the RDF method and the standard features is 1.74\%.

To summarize, we find that the best fitting RDF machine learning redshift using the full 85 input feature list has a decrease in the standard deviation of $29\%$ ($38\%$) for the residual distribution (for the residual distribution scaled by 1/(1+spec\_z) ) when compared to the SDSS DR10 template-ml photometric redshifts. Following \cite{2010A&A...523A..31H} we define the outlier rate as $|\Delta z/(1+$spec\_z$)|>0.15$, and find that the outlier rate of the galaxies in the test sample using the SDSS DR10 template-ml redshift to be $2.32\%$ and using the RDF redshift method the value is reduced to $0.99\%$, which is a 57\% reduction. We note that the hyper-parameters values of the final RDF are \textit{ Number of training examples: 707553, Minimum number of examples per leaf: 6, Number of trees in the forest: 55}. We also note that the most important ranked 1,2,3 features of final RDF are fiberMag\_g, fiberMag\_i and the color dered\_z-dered\_g. We make the datasets and best trained RDF available on the homepage of the lead author upon journal acceptance. 

\section{conclusions}
\label{conclusions}
To exploit large scale photometric surveys the identification of galaxies and measurement of their positions on the sky and in  redshift space is paramount. Very accurate spectroscopic redshifts $z$, can only be measured on a small subset of galaxies due to the integration times required to obtain a reliable measurement.

One may determine a less accurate distance measurement using the photometric properties of the galaxy using either template methods, which encode our parametric knowledge of stellar populations and redshift expansion, or machine learning methods which are non-parametric. In machine learning methods, photometrically measured or derived properties, or `features' are chosen and presented to the machine learning architecture in the hope that the input features identify a good scaling with redshift. 

In this paper we present a study of the effect on the recovered machine learning redshift of the choice of input photometric features. We select and derive 85 easily obtained photometric features of all galaxies with spectroscopic redshifts found in SDSS DR10 CasJobs. We apply very light quality cuts on the recovered galaxies and obtain a sample of 1.9 million galaxies.

We use the machine learning architecture {\rrr Decision Trees combined into Forests (RDF) with Adaboost} \citep[][]{ig,Freund1997119,Drucker:1997:IRU:645526.657132} learning to perform feature importance using the Gini criteria to determine which features produce the most predictive power for redshift estimation, and find that the SDSS $g$ band fiber magnitude is the top ranked, best single feature in $67\%$ of cases. We list the top 1,2,3 features, and show their occurrence fractions after {\rrr randomly exploring} RDF hyper-parameters in Fig. \ref{rankFeat}. Adaboost and RDFs have been used previously to estimate machine learning redshifts but only using standard magnitudes as input features \citep[e.g.][]{2010ApJ...715..823G}. 

We show how the addition of the top 1,2,3 importance ranked features can improve machine learning redshift estimates using the artificial Neural Networks Photoraptor and FaNN. Photoraptor allows a two layer deep network to be trained efficiently, and FaNN implements Cascade2 learning which sequentially adds (multiply connected) hidden neurons to the hidden layer. 

We continue by fixing the Photoraptor machine architecture and present it with the standard and then higher dimensional input features. We find that the recovered machine learning redshift is improved  with the addition of important features. This is in agreement with  \cite{2003LNCS.2859..226T} who show that the addition of extra features such as radii and fluxes improve the machine learning redshift estimation compared with using just magnitudes alone. However note that \cite{2011PASP..123..615S} use more elaborate morphology features and find no real improvements to the machine learning redshifts. In this work we decide which additional features to present to the aNN using the feature importance selection. 

We present the standard and top ranked 1,2 important features to FaNN and allow the machine architecture to vary. While the absolute results are not as competitive as with Photoraptor, the use of the top 1,2 important features does improve the redshift estimate.

To quantify the improvement of the machine learning redshift we measure the residual between the true spectroscopic redshift and the machine learning redshift $\Delta_z$. We measure the mean and the standard deviation of $\Delta_z$ and the fraction of catastrophic `outliers' defined by $|\Delta_z/(1+z)|>0.15$. We find that the addition of the top ranked features to the standard features decreases the standard deviation of the residuals by 13\%, and by 18\% for the residual normalized by 1/(1+$z$), and that the outlier fraction is also decreased by 33\%. We reiterate that this improvement is due to the addition of easily obtainable photometric features. When using FaNN we note that the standard deviation of the residuals decreases by $17\pm8\%$ after marginalizing over the aNN hyper-parameters.

We then show how {\rrr RDFs} can also determine a machine learning redshift \citep[see also][]{2010ApJ...715..823G,tpz}, and document that this technique uses less computational resources, decreases the standard deviation of the residuals \textit{and} lowers the outlier fraction compared with the aNN architectures implemented here. 

We quantify the improvement using {\rrr RDFs} and all 85 input features, with respect to the SDSS DR10 photometric redshift available from CasJobs, and with respect to the two aNN architectures. We find that the standard deviation of the residuals distribution decreases by $\approx29\%$, and by $\approx 38\%$ for the distribution scaled by $1/(1+z)$. We show that the outlier rate of the galaxies in the test sample using the SDSS DR10 photometric redshift is $2.32\%$ and using the RDF method the value decreases to $0.99\%$.

We note that other machine learning architectures are readily available and the use of additional features within these frameworks is an ongoing project. Given the non parametric nature of the machine learning described here, we caution that the results of this analysis are not necessarily easily transported to other surveys or datasets. It would be prudent to perform a similar analysis with different surveys in order to identify their most salient features. However this problem is very tractable. Using {\rrr Decision Trees} and boosting routines, one is able to analyze a dataset of millions of galaxies in just a few hundred seconds using a single core machine. 

\section*{Acknowledgments} 
\label{ack}
{\rrr We would like to thank the referee for useful comments and suggestions which have improved the paper}.
Funding
for the SDSS and SDSS-II has been provided by the Alfred
P. Sloan Foundation, the Participating Institutions, the
National Science Foundation, the U.S. Department of
Energy, the National Aeronautics and Space Administration,
the Japanese Monbukagakusho, the Max Planck
Society, and the Higher Education Funding Council for
England. The SDSS Web Site is http://www.sdss.org/. 

%BIBLIOGRAPHY
\bibliographystyle{mn2e}
\bibliography{photoz}

\end{document}